\documentclass[prl,nopacs,twocolumn,preprintnumbers,amsmath,amssymb]{revtex4}

\usepackage{epsfig}
\usepackage{graphicx}
\usepackage{dcolumn}
\usepackage{color}
\usepackage{bm}

\begin{document}
\title{Limits on electron quality in suspended graphene due to flexural phonons}
\author{E. V. Castro$^*$, H. Ochoa$^*$, M. I. Katsnelson$^{**}$, 
R. V. Gorbachev$^{***}$, D. C. Elias$^{***}$, K. S. Novoselov$^{***}$, 
A. K. Geim$^{***}$, and F. Guinea$^*$}
{\affiliation{$^*$ $^*$Instituto de Ciencia de Materiales de
Madrid (CSIC), Sor Juana In\'es de la Cruz 3, E-28049 Madrid,
Spain. \\ $^{**}$ Radboud University Nijmegen, Institute for Molecules and Materials,
NL-6525 AJ Nijmegen, The Netherlands \\
$^{***}$ School of Physics \& Astronomy and Manchester Centre for
Mesoscience \& Nanotechnology, University of Manchester,
Manchester M13 9PL, UK }

\begin{abstract}
The temperature dependence of the mobility in suspended graphene
samples is investigated. In clean samples, flexural phonons become
the leading scattering mechanism at temperature $T \gtrsim 10\,\,$K, and the
resistivity increases quadratically with $T$. Flexural phonons limit the
intrinsic mobility down to a few $\text{m}^2/\text{Vs}$ at room $T$. 
Their effect can be eliminated by applying strain or placing 
graphene on a substrate.
\end{abstract}
\maketitle

{\em Introduction.---}The properties of isolated graphene continue to
attract enormous interest due to both its exotic electronic properties
\cite{NGPNG09} and realistic prospects of various applications \cite{AGSci09}.
It has been found that the intrinsic mobility $\mu$ of charge carriers in
graphene can exceed $20\,\,\text{m}^2/\text{Vs}$ at room temperature $T$ 
\cite{Metal08,CXI+07}, which is the absolute record. So far, such high values
have not been achieved experimentally, because extrinsic scatterers
limit $\mu$. The highest $\mu$ was reported in suspended devices 
\cite{DSBA08,Betal08} and could reach $\sim 12\,\, \text{m}^2/\text{Vs}$ at
$240\,\,\text{K}$ \cite{BSHSK08}. This however disagrees with the data of
Ref.~\cite{DSBA08} where similar samples exhibited room $T$ $\mu$ close to
$\sim 1\,\,\text{m}^2/\text{Vs}$, the value that is routinely achievable 
for graphene on a substrate.

In this Letter, we show that flexural phonons (FP) are an important scattering
mechanism in suspended graphene and the likely origin of the above disagreement,
and their contribution should be suppressed to allow ultra high $\mu$. 
Generally, electron-phonon scattering in graphene is expected to be weak due
to very high phonon frequencies \cite{HdSacph07}. However, in suspended
thin membranes, out of plane vibrations lead
to a new class of low energy phonons, the flexural
branch \cite{LL59,N89}. In an ideal flat suspended membrane symmetry
arguments show that electrons can only be scattered by two
FP simultaneously \cite{Metal08,MO08}. As a result
the resistivity due to FP rises rapidly at high $T$ where it can be
described as elastic scattering by thermally excited intrinsic
ripples \cite{KG08}.

We analyze here the contribution of FP to
the resistivity, and present
experimental results which strongly support the suggestion that
FP are a major source of electron scattering in
suspended graphene. This intrinsic limitation to the achievable
conductivity of graphene at room $T$ can be relaxed by
applying tension, which modifies both the phonons and their
coupling to charge carriers.

{\em Model.---}Graphene is a two dimensional membrane, whose
elastic properties are well described by the free energy
\cite{LL59,N89}:
\begin{align}
{\cal F} &\equiv \frac{1}{2} \kappa \int dx dy \left( \nabla^2 h
\right)^2 + \frac{1}{2} \int dx dy \left( \lambda u_{ii}^2 + 2 \mu
u_{ij}^2 \right) \label{free_energy}
\end{align}
where $\kappa$ is the bending rigidity, $\lambda$ and $\mu$ are
Lam\'e coefficients, $h$ is the displacement in the out of plane
direction, and $u_{ij} = 1/2
\left[ \partial_i u_j + \partial_j u_i + ( \partial_i h ) (
\partial_j h ) \right]$ is the strain tensor. Summation over indices in
Eq.~\eqref{free_energy} is implied. Typical parameters for
graphene \cite{ZKF09,Hone08,KSY01} are $\kappa \approx 1$~eV, and $\mu
\approx 3 \lambda \approx 9$~eV~\AA$^{-2}$. The density is
$\rho = 7.6 \times 10^{-7}$~Kg/m$^2$. The velocities of the
longitudinal and transverse phonons obtained from
Eq.~\eqref{free_energy} are $v_L = \sqrt{\frac{\lambda+2\mu}{\rho}}
\approx 2.1 \times 10^4$~m/s and $v_T = \sqrt{\frac{\mu}{\rho}}
\approx 1.4 \times 10^4$~m/s. The FP show the
dispersion
\begin{align}
\omega_{\vec{\bf q}}^F &= \alpha \left| \vec{\bf q} \right|^2
\label{flexural}
\end{align}
with $\alpha = \sqrt{\frac{\kappa}{\rho}} \approx 4.6 \times
10^{-7}$~m$^2$/s.

Suspended graphene can be under tension, either due to the
electrostatic force arising from the gate, or as a result of 
microfabrication. Let us assume that there are slowly varying in
plane stresses, $u_{ij} ( \vec{\bf r} )$, which change little on the
scale of the Fermi wavelength, $k_F^{-1}$, which is the relevant
length for the calculation of the carrier resistivity. Then, the
dispersion in Eq.~\eqref{flexural} is changed into:
\begin{align}
\omega_{\vec{\bf q}}^F ( \vec{\bf r} )&= | \vec{\bf q} |
\sqrt{\frac{\kappa}{\rho} | \vec{\bf q} |^2 + \frac{\lambda}{\rho}
u_{ii} ( \vec{\bf r} ) + \frac{2 \mu}{\rho} u_{ij} ( \vec{\bf r} )
\frac{q_i q_j}{| \vec{\bf q} |^2}} \label{flexuralstr}
\end{align}
The dispersion becomes anisotropic. For small wavevectors, the
dispersion is linear, with a velocity which scales as 
$\sqrt{ \bar u }$, where $\bar u$ is strain.

The coupling between electrons and long wavelength phonons can be
written in terms of the strain tensor. On symmetry grounds, we can
define a scalar potential and a vector potential which change the
effective Dirac equation which describes the electronic
states \cite{SA02b,M07,NGPNG09,VKG10}:
\begin{align}
V ( \vec{\bf r}  ) &= g_0 \left[ u_{xx} ( \vec{\bf r} ) + u_{yy} (
\vec{\bf
r} ) \right] \nonumber \\
\vec{\bf A} ( \vec{\bf r} ) &= \frac{\beta}{a} \left\{ \frac{1}{2}
\left[ u_{xx} ( \vec{\bf r} ) - u_{yy} ( \vec{\bf r} ) \right] ,
u_{xy} ( \vec{\bf r} ) \right\} \label{fields}
\end{align}
where $g_0 \approx 20 - 30$~eV is the bare deformation potential
\cite{SA02b},
$a \approx 1.4$~\AA\, is the distance between nearest carbon atoms,
$\beta = - \partial \log ( t ) /
\partial \log ( a ) \approx 2-3$ \cite{HKSS88}, and $t \approx 3$~eV is the
hopping between electrons in nearest carbon $\pi$ orbitals.

Linearizing Eq.~\eqref{free_energy} and expressing the atomic
displacements in terms of phonon creation and destruction operators,
and using Eq.~\eqref{fields} and the Dirac Hamiltonian for
graphene~\cite{NGPNG09} we can write the full
expressions for the coupling of charge carriers to longitudinal,
transverse and FP, without and with preexisting
strains.

{\em  Calculation of the resistivity.---}We assume that the phonon
energies are much less than the Fermi energy, so that the electron
is scattered between states at the Fermi surface. After some
algebra, the scattering rate due to FP, including a
constant strain, $\bar{u}$, is
\begin{widetext}
\begin{align}
\frac{1}{\tau_F} &= \frac{1}{ 32 \pi^3 \rho^2 v_F k_F}
\int_0^{2k_F} d K \frac{[D(K)]^2K^2}{\sqrt{k_F^2 - K^2/4}}
\int_0^\infty d q \frac{q^3 n_q}{\omega_q}
\int_{|K-q|}^{|K+q|} d Q \frac{Q^3 (n_Q + 1)}{\omega_Q \sqrt{K^2 q^2 - ( K^2 + q^2 - Q^2 )/4}}
\label{rate}
\end{align}
\end{widetext}
where $v_F$ is the Fermi velocity and $k_F$ the Fermi momentum,
$[D(K)]^2 = [g(K)]^2[1 - K^2/(2k_F)^2]+ ( \beta \hbar v_F )^2 /(4  a^2)$ is the
generalized deformation potential, including the contribution of the
screened scalar potential $g(K) = g_0/\varepsilon (K) $ and gauge potential,
$\omega_q = \sqrt{\alpha^2 q^4 + \bar{u} v_L^2 q^2}$ is the phonon
dispersion, Eq.~\eqref{flexural} in the isotropic approximation, and $n_q$ is
the Bose-Einstein distribution function. The diagram described in this
calculation, and the variables $K = | \vec{\bf K} |$, $q = | \vec{\bf q} |$
and $Q = | \vec{\bf Q} |$ are shown in Fig.~\ref{diagram}. The static
dielectric function is $\varepsilon(K) = 1 + e^2 N(k_F)/(2 \epsilon_0 K)$,
where $N ( k_F ) = ( 2 k_F ) / ( \pi \hbar v_F )$ is the density of states.
At $k_F$ the screened scalar potential
$g\equiv g_0/\varepsilon(k_F) \approx g_0/8 \approx 3\,\text{eV}$ is in good
agreement with \emph{ab initio} calculations \cite{CJS10}.

\begin{figure}
\includegraphics[width=0.9\columnwidth]{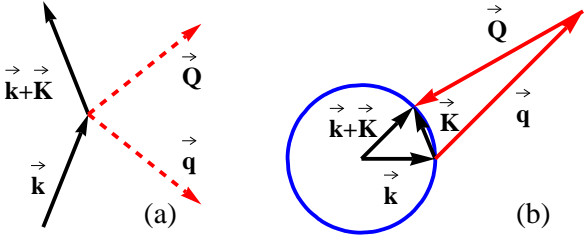}
\caption[fig]{(color online). a) Two phonon diagram which
describes electron scattering. b) Kinematics of the process, and
variables used in Eq.~(\ref{rate}). The circle denotes the Fermi
surface.} \label{diagram}
\end{figure}

The relevant phonons which contribute to the resistivity are those
of momenta $| \vec{\bf q}  | \gtrsim 2 k_F$.  This scale allows us
to define the Bloch-Gr\"uneisen temperature, $k_B T_{BG} = \hbar
\omega_{2 k_F}$. Neglecting first the effect of strain, we find:
\begin{align}
T_{BG}^{L} &= 57 \sqrt{n}\,\, {\rm K} \nonumber \\
T_{BG}^{T} &= 38 \sqrt{n}\,\, {\rm K} \nonumber \\
T_{BG}^{F} &= 0.1 n\,\, {\rm K} \label{BG}
\end{align}
respectively for in-plane longitudinal ($L$) and transverse ($T$)
and for FP ($F$), where the temperature is in Kelvin
and the electron density $n$ is expressed in $10^{12}$~cm$^{-2}$.
 Close to
room $T$ we are in the regime $T \gg
T_{BG}^{i=L,T,F}$ for all concentrations of interest. The
corresponding temperature for FP in the presence of a
uniaxial strain, $\bar{u}$ is $T_{BG} = 28 \sqrt{\bar{u} n}$~K.
Our focus here is on the experimentally relevant high--$T$ regime.

In systems with strain, the phonon dispersion relation, Eq.~(\ref{flexural}),
shows a crossover between a regime dominated by the to another where
the strain becomes irrelevant, at $q^* \approx v_L \sqrt{\bar{u}} / \alpha$.
The range of integration over the phonon momenta in Eq.~(\ref{rate}) is
limited by $\hbar \omega_q \lesssim k_B T$, and $k_F \lesssim q$. 
In addition the theory has a natural infrared cutoff with a characteristic
momentum $q_c$ below which the anharmonic effects become important 
\cite{ZRF+10}.
Defining $q_T$ as $\hbar \omega_{q_T} = k_B T$, the scattering rate in
Eq.~(\ref{rate}) shows three regimes in which (i) strain is irrelevant and
$\max(q^*,q_c) \ll k_F$, (ii) strain is small and relevant phonons combine  
linear and quadratic spectrum for $\max(k_F,q_c) \lesssim q^* \lesssim q_T$,
(iii) strain is high and determines the scattering rate for $q_T \ll q^*$. 
We finally obtain:
\begin{equation}
\frac{1}{\tau_F} \approx 
\begin{cases}
\frac{D^2 (k_B T)^2 }{64 \pi \hbar^2 \kappa^2 v_F k_F} 
\ln(\frac{k_B T}{\hbar \omega_{c}})
&\max(q^*,q_c) \ll k_F \ll q_T\\
\frac{D^2 (k_B T)^2 k_F}{32 \pi \hbar^2 \rho  \kappa  v_F v_L^2 \bar{u}} &
\max(k_F,q_c) \ll q^* \ll q_T \\
\frac{6 \zeta(3) D^2 (k_B T)^4 k_F}{16 \pi \hbar^4 \rho^2 v_F v_L^{6} \bar{u}^{3}}
&k_F \ll q_T \ll q^*
\end{cases},
\label{temperature}
\end{equation}
where $D^2 = g^2/2 +( \beta \hbar v_F )^2 /(4  a^2)$, and the infrared cutoff
$\hbar \omega_c$ is related to $\max(q^*,q_c)$.
For comparison we give also the contribution from in-plane phonons,
\begin{equation}
 \frac{1}{\tau_{L,T}} \approx  \left[ \frac{g^2}{v_L^2}  +
\frac{\beta^2 \hbar^2 v_F^2}{4 a^2} \left( \frac{1}{v_L^2} +
\frac{1}{v_T^2} \right) \right] \frac{k_F k_B T}{2 \rho \hbar^2 v_F}.
\label{eq:tauinplane}
\end{equation}

The $T$ dependence of the scattering due to FP
is more pronounced than that due to in-plane phonons, and
it dominates at high enough $T$. In the limit of irrelevant
strains, $\max(q^*,q_c) \ll k_F$,
the crossover temperature is
\begin{align}
\frac{\tau_{L,T} ( T^* )}{\tau_F ( T^* )}  &= 1
\Rightarrow T^* ( {\rm K} ) \approx 57 \times n ( 10^{12} {\rm cm}^{-2} ).
\label{cross1}
\end{align}
When $T^* \lesssim T^{L,F}_{BG}$ this crossover does not occur and scattering
by FP dominates also at low temperatures.
At finite strain $\max(k_F,q_c) \ll q^* \ll q_T$ we obtain
\begin{align}
\frac{\tau_{L,T} ( T^* )}{\tau_F ( T^* )}  &= 1 \Rightarrow T^* ( {\rm K} ) \approx  10^6   \bar{u}.
\label{cross2}
\end{align}

In the absence of strains, the crossover shown in Eq.~\eqref{cross1} implies
that the room $T$ mobility is limited by FP for
densities below $10^{13}$cm$^{-2}$. Strains reduce significantly the effect of
FP, so that, in the presence of strain, the mobility is
determined by the scattering by in-plane phonons, see Eq.~\eqref{cross2}.

The contribution to the resistivity from the different phonon modes can be
written, using the expressions for the scattering rate as
\begin{align}
\rho_i ( n , T , \bar{u} ) &= \frac{2}{e^2 v_F^2 N ( k_F ) \tau_i ( n , T , \bar{u} )}
\label{resistivity}
\end{align}
where the index $i$ label the phonon mode. Results for the resistivity in
different regimes are shown in Fig.~\ref{comp}.

\begin{figure}
\includegraphics[width=0.9\columnwidth]{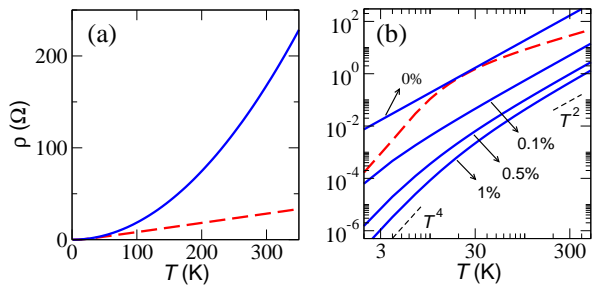}
\caption[fig]{(Color online). (a) Contribution to the resistivity
from flexural phonons (blue full line)
and from in-plane phonons (red dashed line). (b) Resistivity for different
strain. The in-plane contribution (broken red line) shows
a crossover from a low to a high--$T$ regime.  In both cases, the
electronic concentration is $n = 10^{12}$cm$^{-2}$.}
\label{comp}
\end{figure}

The results of Eqs.~\eqref{temperature} and~\eqref{resistivity}
can be extended to bilayer graphene. The main differences are: (i)
the kinematics of the two phonon scattering are the same as in
Fig.~\ref{diagram}, except that the overlap between the electronic
states $| \vec{\bf k} \rangle$ and $| \vec{\bf k} + \vec{\bf K}
\rangle$ is modified; (ii) the density of states is constant
$N(\epsilon) \equiv N_0 = t_\perp /(\pi \hbar^2 v_F^2)$, and the
screened scalar potential $g$ is replaced by $(2\epsilon_0 g_0 k_F
) / [ e^2 N ( \epsilon_F ) ] = ( 2\pi\epsilon_0 g_0 k_F \hbar^2
v_F^2 ) / ( e^2 t_\perp )$, where $t_\perp$ is the hopping between
layers, and $e$ is the electric charge;  (iii) the Fermi velocity,
which determines the coupling to the gauge potential is $2 v_F^2
\hbar k_F / t_\perp$. The Fermi velocity and the density of states
also change the expression for the resistivity,
Eq.~\eqref{resistivity}.

{\em Experimental results.---}We have fabricated two-terminal suspended devices
following the procedures introduced in Refs.~\cite{DSBA08,Betal08}. Typical 
changes in the resistance $R$ as a function of the gate induced concentration 
$n$ are shown in Fig. 3a. The as-fabricated devices exhibited 
$\mu \sim 1\,\, \text{m}^2/\text{Vs}$ but, after their in situ annealing by 
electric current, $\mu$ could reach above $100\,\, \text{m}^2/\text{Vs}$ at 
low $T$. 
To find $\mu$, we have used the standard expression $R = R_0 +(l/w)(1/ne\mu)$ 
where $R_0$ describes the contact resistance plus the effect of neutral 
scatterers, and both $R_0$ and $\mu$ are assumed $n$-independent 
\cite{Metal08,CXI+07}. Our devices had the length 
$l \approx 1-2\,\,\mu \text{m}$ and 
the channel width $w$ of $2-4\,\,\mu \text{m}$ (see the inset in Fig. 3b). 
At $T > 100\,\,\text{K}$, the above expression describes well the functional 
form of the experimental curves, yielding a constant $\mu$ over the whole 
range of accessible $n$, if we allow $R_0$ to be different for electrons and 
holes. This is expected because of an $n-p$ barrier that appears 
in the regime of electron doping due to our $p$-doping contacts 
\cite{DSBA08,Betal08}. At $T < 100\,\,\text{K}$, the range of $n$ over which 
the expression fits the data rapidly narrows. Below $20\,\,\text{K}$, we can 
use it only for $n < \pm 10^{10}\,\,\text{cm}^{-2}$ because at higher $n$ we 
enter into the ballistic regime (the mean free path, proportional to 
$ \mu n^{1/2}$, becomes comparable to $l$). In the ballistic regime, 
graphene's conductivity $\sigma$ is no longer proportional to 
$n$ \cite{DSBA08,Betal08} and the use of $\mu$ as a transport parameter has 
no sense. To make sure that $\mu$ extracted over the narrow range of $n$ is 
also correct, we have crosschecked the found $\mu$ against quantum mobilities 
inferred from the onset of Shubnikov-de Haas oscillations 
\cite{DSBA08,Betal08}. For all our devices with $\mu$ ranging from 
$\sim 1-100\,\,\text{m}^2/\text{Vs}$, we find good agreement between transport and 
quantum mobilities at liquid-helium $T$. Fig. 3b shows the $T$ dependence of 
$\mu$. It is well described by the quadratic dependence 
$1/\mu =1/\mu(T\rightarrow 0) + \gamma T^2$. Surprisingly, we find the 
coefficient $\gamma$ to vary by a factor of $\sim 2$ for different devices, 
which is unexpected for an intrinsic phonon contribution. 
Such variations are 
however expected if strain modifies electron-phonon scattering as 
discussed below.
 Note that $\mu$ falls down to $4-7\,\,\text{m}^2/\text{Vs}$ at 
$200\,\,\text{K}$ (see Fig. 3b) and the extrapolation to room $T$ yields 
$\mu$ of only $2-3\,\,\text{m}^2/\text{Vs}$, which is significantly lower 
than the values reported in Ref.~\cite{Betal08} but in agreement with 
Ref.~\cite{DSBA08}.

\begin{figure}
\includegraphics[width=0.9\columnwidth]{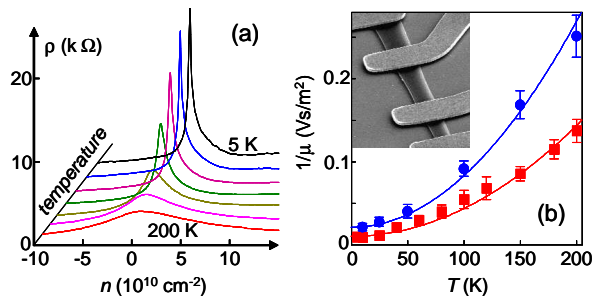}
\caption[fig]{(color online). (a) Electron transport in suspended graphene. 
Graphene resistivity $\varrho = R (w/l)$ as a function of gate-induced 
concentration $n$ for $T = 5,$ 10, 25, 50, 100, 150 and 200~K. (b) Examples 
of $\mu(T)$. The $T$ range was limited by broadening of the peak beyond the 
accessible range of $n$. The inset shows a scanning electron micrograph of one 
of our suspended device. The darker nearly vertical stripe is graphene 
suspended below Au contacts. The scale is given by graphene width of 
about $ 1\,\, \mu \text{m}$ for this particular device.}
 \label{exp}
\end{figure}

{\em Discussion.---}The density independent $\mu$ indicates that experiments are
in the non-strained regime, where $1/\tau_F \sim T^2 / k_F$ and
$\rho_F \sim T^2 / n$. Here FP completely dominate
and the coefficient $\gamma$ defined above is given by
$\gamma \approx \frac{D^2 k_B^2 }{64 \pi e \hbar \kappa^2 v_F^2} 
\ln(\frac{k_B T}{\hbar \omega_{c}})$, where the infrared cutoff 
is the only free parameter \cite{foot}.
Experiment gives $\gamma \approx 6.19\times 10^{-6}\,\,\text{Vs}/(\text{mK})^2$
for the sample with lower mobility and 
$\gamma \approx 3.32\times 10^{-6}\,\,\text{Vs}/(\text{mK})^2$
 for the higher mobility one. Neglecting the logarithmic correction of order 
unity, the analytic expression gives 
$\gamma \approx 3\times 10^{-6}\,\,\text{Vs}/(\text{mK})^2$ without 
adjustable parameters.

The difference between the two samples may be understood as due
to a different cutoff under the logarithm due to strain. 
In non-strained samples
there is a natural momentum cutoff $q_c \approx 0.1\,\,\text{\AA}^{-1}$ 
below which the harmonic approximation breaks down \cite{ZRF+10}.
Strain increases the validity of harmonic approximation, making $q_c$
strain dependent, thus explaining different cutoff at different
strain. A rough estimate of the expected strains is obtained 
by comparing $q_c \approx 0.1\,\,\text{\AA}^{-1}$ with 
$q^*=v_L \sqrt{\bar u}/\alpha$, which gives $\bar u \sim  10^{-4} - 10^{-3}$.
Such small strain can be present even in slacked samples (where strain
induced by gate and $T$ is negligible) due to, for example, 
the initial strain induced by the substrate and remaining unrelaxed
under and near metal contacts. A complete theory would require
the treatment of anharmonic effects, which is beyond the scope of
the present work. The data in \cite{BSHSK08} show higher mobilities than
those in Fig.~\ref{exp}. A fit to this data using
Eq.~\eqref{temperature} suggests the sample being under strain.

{\em Conclusions.---}The experimental and theoretical results
presented here suggest that FP are the main mechanism
which limits the resistivity in suspended graphene samples, at
temperatures above $10\,\,\text{K}$. Scattering by FP involves
two modes, leading to a $T^2$ dependence at high temperatures, with 
mobility independent of carrier concentration. These
results agree qualitatively with classical theory assuming elastic
scattering by static thermally excited ripples \cite{KG08}.
Quantitatively, one of our main results is that in devices with
negligible strain the mobility
does not exceed values of the order of $1\,\,$m$^2$V$^{-1}$s$^{-1}$ 
at room $T$, that is, FP 
restrict the electron mobility to values typical for exfoliated
graphene on a substrate.

The dispersion of FP changes from quadratic to
linear when the sample is under tension. As a result, the
influence of FP on the transport properties is
suppressed. The $T$ dependence of the mobility remains
quadratic, but it decreases linearly with the carrier
concentration. Importantly, applying rather weak strains may be enough
to increase dramatically the mobility in freely suspended samples
at room $T$.

A very recent theory work \cite{MvO10} has also addressed the role of
FP on electron transport. Insofar as the two analysis partially  overlap, 
the results are in agreement.

{\em Acknowledgments.---}Useful discussions with Eros Mariani are
gratefully acknowledged. We acknowledge financial support from
MICINN (Spain) through grants FIS2008-00124 and CONSOLIDER
CSD2007-00010, and from the Comunidad de Madrid, through
NANOBIOMAG. MIK acknowledges a financial support from FOM (The
Netherlands).


\begin{thebibliography}{23}
\bibitem[{foo()}]{foot}
\bibinfo{note}{We fixed $g \approx 3\,\,\text{eV}$ and $\beta \approx 3$. These
  values are not only in agreement with theoretical predictions
  \cite{HKSS88,CJS10}, they also perfectly reproduce the experimental data in
  Ref.~\cite{CXI+07} using Eqs.~\eqref{eq:tauinplane} and \eqref{resistivity}.}
\end{thebibliography}
\end{document}